\documentclass[aps,prb,twocolumn,amsmath,amssymb,showpacs]{revtex4-1}
\usepackage{graphicx}
\usepackage{dcolumn}
\usepackage{bm}
\usepackage{color}
\newcommand{\comment}[1]{}
\newcommand{\cG}{{\cal G}}

\begin{document}

\title{Spin thermoelectrics in a disordered Fermi gas}

\author{J. Borge$^{a}$,  C. Gorini$^{b}$,  R. Raimondi$^a$  }
\affiliation{
$^a$CNISM and Dipartimento  di Fisica "E. Amaldi", via della Vasca Navale 84, Universit\`a  Roma Tre, 00146 Roma, Italy\\
$^b$Institut f\"ur Physik, Universit\"at Augsburg, 86135 Augsburg, Germany
}

\date{\today}

\begin{abstract}
We study the connection between the spin-heat and spin-charge response 
in a disordered Fermi gas with spin-orbit coupling.
It is shown that the ratio between the above responses can be expressed 
as the thermopower $S=-(\pi k_B)^2T\sigma'/3e\sigma$ times a number $R_s$ 
which depends on the strength and type of the spin-orbit couplings considered.
The general results are illustrated by examining different two-dimensional
electron or hole systems with different and competing spin-orbit mechanisms,
and we conclude that a metallic system could prove much more efficient
as a heat-to-spin than as a heat-to-charge converter. 
\end{abstract}

\maketitle

\section{Introduction}
The moving carriers in a metallic system, electrons or holes, transport both electric charge and heat.
This gives rise to a number of thermoelectric effects as well as a deep connection between
thermal and electrical conductivities.  A well known example is the Wiedemann-Franz law, which states that the ratio
of the thermal to the electrical conductivity is the temperature times a universal number,
the Lorenz number ${\cal L}=\pi^2k_B^2/3e^2$, where $k_B$ and $e$ are the Boltzmann constant and the unit charge.
From the theoretical standpoint the validity of the above law relies on the single-particle description of transport, 
on the Fermi statistics of carriers and on the assumption of purely elastic scattering \cite{chester1961,jonson1980}.
When electron-electron interaction is present as in a Fermi liquid, 
this law still holds provided the quasiparticles do not exchange energy during collisions. 
At low temperatures the combination of electron-electron interaction and disorder may change this picture
\cite{langer1962,castellani1987,livanov1991,raimondi2004,niven2005,catelani2005,michaeli2009}.
Additionally, a magnetic field affects both thermal and electrical transport yielding 
both galvanomagnetic and thermomagnetic effects \cite{sondheimer1948}.
The above situation gets even more complicated when a third quantity transported by
the carriers -- the spin -- is connected to the previous two by spin-orbit (SO) coupling.
On the bright side, such a connection also opens up a plethora of new possibilities
related to the manipulation of the additional spin degrees of freedom.
This is testified by the recent rapid development of spintronics \cite{zutic2004,awschalom2007}
and spin caloritronics \cite{bauer2012}.
A fundamental goal of spintronics is the achievement of all-electrical control
of the carriers' spin, made possible by SO coupling
as exemplified by the spin Hall effect \cite{dyakonov1971,hirsch1999,murakami2003,sinova2004,vignale2010}. 
Similarly, an important goal of spin caloritronics is the manipulation of 
the spin degrees of freedom via thermal gradients \cite{wang2010,ma2010,nunner2011,slachter2011,scharf2011},
particularly relevant when energy efficiency issues are considered \cite{bauer2012}.
In this context a noteworthy phenomenon is the spin Seebeck effect:
a spin current thermally generated in a (metallic or insulating) ferromagnet
is injected into a normal metal and there, via the inverse spin Hall effect,
it generates an observable voltage drop in the direction orthogonal 
to the applied thermal gradient \cite{uchida2008, uchida2010, jaworski2010}.
In this much studied case phonons and magnons play the leading roles \cite{jaworski2011, uchida2012, adachi2012}.
There are on the other hand only few works on thermo-spin transport
due to the charge carriers' dynamics \cite{ma2010, nunner2011}, and we wish to address this issue
considering disordered Fermi gases with SO coupling.
We will see that a general relation between the spin-heat and spin-charge response
of such systems can be obtained, with the same range of applicability 
of the Wiedemann-Franz law.  Moreover, we will discuss the particular
case of the thermo-spin Hall effect -- the generation of a spin current
transverse to a thermal gradient, also called the spin Nernst effect.  
In so doing we will show that a simple relation connects 
the spin thermopower -- the ratio between the spin response to a thermal gradient 
and that to an electric field -- to the standard electric thermopower, 
and that the former can be strongly enhanced by the interplay between different SO coupling mechanisms.

Let us start with some basic phenomenological considerations
along the lines of Refs.~\onlinecite{dyakonov2007, schwab2010},
and consider the bare-bones situation of an inversion symmetric,
homogeneous and non-ferromagnetic material in the absence of magnetic fields.
A particle current $j_x$ can be driven either by an electric field
or by a temperature gradient, and
within the standard semiclassical approach one writes \cite{aschcroftbook}
\begin{equation}
\label{phenom1}
 j_x = L_{11}E_x + L_{12}(-\nabla_xT) = \sigma E_x - e{\cal L}T\sigma' (-\nabla_xT).
\end{equation}
Here $\sigma=-2eN_0D$ is the Drude conductivity up to a charge $-e$, 
with $N_0$ the density of states at the Fermi energy and $D$ the diffusion constant,
and $\sigma'=\partial_{\mu}\sigma$, $\mu$ being the chemical potential.  
The ratio $S\equiv L_{12}/L_{11}$ is the electric thermopower.
In the present simple case the connection between spin and particle currents
due to SO coupling reads \cite{dyakonov2007}
\begin{equation}
\label{phenom2}
j^z_y = -\gamma j_x = L^s_{11}E_x + L^s_{12}(-\nabla_xT). 
\end{equation}
Here $j^z_y$ is the $z$-polarized spin current flowing in the $y$ direction
arising in response to the particle current $j_x$, 
and $\gamma\ll1$ is a dimensionless SO coupling constant.
As an immediate consequence of Eqs.~(\ref{phenom1}) and (\ref{phenom2}),
the spin thermopower $S_s\equiv L^s_{12}/L^s_{11}$ is equal to $S$,
since the SO coupling constant $\gamma$ does not depend on the sources of a given particle current.
Eq.~\eqref{phenom2} breaks down in the absence of inversion symmetry, 
and in order to see how the above simple result is modified
in a general situation, and to study its dependence on competing
SO coupling mechanisms, we will move on to a microscopic treatment.

The paper is organized as follows.  The formalism is introduced in Sec.~\ref{sec_basic}
and put to use in Sec.~\ref{sec_spin_mott} to obtain the general formula for the spin-thermopower $S_s$.
The latter appears as the spin equivalent of Mott's formula for the electric thermopower.
In order to lend concreteness to the presentation, the derivation of $S_s$ is done
using the linear Rashba model as a template.
In Sec.~\ref{sec_appl} we apply our formula to a series of different systems
and discuss its experimental relevance, before concluding in Sec.~\ref{sec_conclusions}.
A number of technical details regarding the Matsubara technique are presented in the Appendix.


\section{The basic equations}
\label{sec_basic}

Though our treatment is independent of dimensions (2D or 3D), in order to fix things 
we consider a disordered 2D Fermi gas in the $x$-$y$ plane described by the Hamiltonian
\begin{equation}
\label{eq1}
H=\frac{p^2}{2m}+V({\bf x})+H_{\rm so},
\end{equation}
with ${\bf p}$ the 2D momentum and $V({\bf x})$ the impurity potential.
For the latter we assume the standard white noise disorder model
and evaluate the impurity average in the Born approximation,
$\langle V({\bf x})V({\bf x'})\rangle=(2\pi N_0 \tau)^{-1}\delta ({\bf x}-{\bf x'})$,
with $N_0=m/(2\pi\hbar^2)$ and $\tau$ the elastic scattering time.
The SO term $H_{\rm so}$ will have different forms in the various cases considered below. 
In the (linear) Rashba case it reads
\begin{equation}
\label{eq2}
H_{so}=\alpha {\boldsymbol \sigma} \cdot  {\bf p}\times {\bf {\hat e}}_z,
\end{equation}
with $\alpha$ a coupling constant.
We assume the metallic regime and weak SO coupling conditions,
$\epsilon_F \gg \hbar/\tau, \Delta_{\rm so} $.  
Here $\epsilon_F$ is the Fermi energy in the absence of disorder and SO interaction
and $\Delta_{\rm so}$ is the SO splitting due to $H_{\rm so}$.
From now on $\hbar,k_B =1$.
The $a$-polarized spin current flowing in the $k$-direction due to a generic thermal gradient is
\begin{equation}
\label{eq3}
j_k^a =\sum_l\,\left[{\rm N}_{\rm sh}\right]^a_{kl} \left(-\partial_l T \right),
\end{equation}
where ${\rm N}_{\rm sh}$ is the spin-heat response tensor.
Following Ref.~\onlinecite{niven2005} the latter is given in terms of the imaginary spin current-heat current kernel
\begin{equation}
\label{kappa_sh}
\left[{\rm N}_{\rm sh}\right]^a_{kl}T= \lim_{\Omega \rightarrow 0}
\left\{
\frac{\left[Q_{\rm sh}({\rm i}\Omega_{\nu})\right]^a_{kl}}{\Omega_{\nu}}
\right\}_{{\rm i}\Omega_{\nu}\rightarrow \Omega^R, \  \Omega^R=\Omega +{\rm i}0^+}.
\end{equation}
The spin current operator is given by the standard definition $j^a_k=(1/2) \{ v_k, s^a\}$, 
$v_k$ and $s^a$ being the velocity and spin operators.
Notice that the particle (charge) current operator is $(-e)j_k=(-e)v_k$. 
The heat current in the Matsubara representation reads
\begin{equation}
\label{eq5}
j^h_k ({\bf p}, \epsilon_n, \epsilon_n +\Omega_{\nu})= {\rm i} \epsilon_{n+\nu /2} j_k,
\end{equation}
with $\epsilon_n=\pi T(2n+1), \Omega_{\nu}=2\pi T \nu$, and $\epsilon_{n+\nu/2}=\epsilon_n+\Omega_{\nu}/2$.
The specific form of $v_k$ depends on the choice of the SO Hamiltonian.
For instance in the Rashba case, Eq.(\ref{eq2}), we have $v_{x,y}=p_{x,y}/m \mp \alpha \sigma^{y,x}$.
By using the Kubo formula the response kernel is given by
\begin{equation}
\label{Q_sh}
\left[Q_{\rm sh}\right]^a_{kl}({\rm i}\Omega_{\nu})
=T\sum_{\epsilon_n ,\bf p} {\rm i} \epsilon_{n+\nu/2} 
{\rm Tr}\left[ j^a_k \cG_n j_l\cG_{n+\nu}\right],
\end{equation}
where the Matsubara Green functions $\cG_n=\cG({\bf p},\epsilon_n)$, $\cG_{n+\nu}=\cG({\bf p},\epsilon_n + \Omega_{\nu})$
are matrices in spin space $\cG_n=\cG^0_n + \sum_a\; \cG^a_n \sigma^a$.
Analogously, the spin-charge response kernel can be written as
\begin{equation}
\label{kernel_spin_charge}
\left[Q_{\rm sc}\right]^a_{kl}({\rm i}\Omega_{\nu})
=
-e T
\sum_{\epsilon_n,\bf p}{\rm Tr}\left[ j^a_k \cG_n j_l\cG_{n+\nu}\right],
\end{equation}
leading to the spin-charge (particle) conductivity
\begin{equation}
\label{sigma_spin_charge}
\left[{ \sigma}_{\rm sc}\right]^a_{kl}= \lim_{\Omega \rightarrow 0}
\left\{
\frac{\left[Q_{\rm sc}({\rm i}\Omega_{\nu})\right]^a_{kl}}{\Omega_{\nu}}
\right\}_{{\rm i}\Omega_{\nu}\rightarrow \Omega^R, \  \Omega^R=\Omega +{\rm i}0^+}.
\end{equation}


\section{The spin equivalent of Mott's formula}
\label{sec_spin_mott}

Although our treatment is general, to illustrate the procedure we take the Rashba case as an example.
The average over disorder is evaluated in the Born approximation and leads to a self-energy
\begin{equation}
\label{self_energy}
\Sigma (\epsilon_n ) =\frac{1}{2\pi N_0 \tau}\sum_{\bf p} \cG_n =-\frac{{\rm i}}{2\tau}{\rm sgn } (\epsilon_n),
\end{equation}
which is diagonal in spin space. As it can be seen from Eq.(\ref{eq2}) for the Rashba case,
the off-diagonal terms in spin space of the Green function are odd in the momentum dependence and vanish upon integration. This remains valid also for other spin-orbit interaction terms as long as the Hamiltonian is time-reversal invariant.

To compute the thermo-spin Hall effect.  i.e.   the $z$-polarized spin current flowing along $y$
generated by a thermal gradient along $x$, we need
the response kernel $\left[Q_{\rm sh}\right]^z_{yx}\equiv Q^{sH}$, which  reads
\begin{equation}
\label{eq6}
Q^{sH}({\rm i}\Omega_{\nu})=T\sum_{\epsilon_n}\sum_{\bf p} {\rm i} \epsilon_{n+\nu/2} 
{\rm Tr}\left[ j^z_y \cG_n j_x\cG_{n+\nu}\right],
\end{equation}
with $\cG^0_n=(\cG_{n,+}+\cG_{n,-})/2$ and $\cG^a_n=(\hat{\bf p}\times\hat{\bf e}_z)^a\left(\cG_{n,+}-\cG_{n,-}\right)/2$, whereas
\begin{equation}
\label{cGrashba}
\cG_{n,\pm} = \left[{\rm i}\epsilon_n+\mu-\frac{p^2}{2m}\mp\alpha p+\frac{{\rm i}}{2\tau}{\rm sgn}(\epsilon_n)\right]^{-1},
\end{equation}
 $\mu$ being the chemical potential. 
 
Notice that the analytic properties of the Green functions are determined by the sign of the imaginary frequency,
therefore when performing the momentum integral in Eq.(\ref{eq6}) one obtains a non-zero result only if the frequencies
$\epsilon_n+\Omega_{\nu}$ and $\epsilon_n$ have opposite signs, which means that
$\epsilon_n$ is restricted to the range $-\Omega_{\nu} < \epsilon_n <0$.
Exploiting that the external frequency is going to zero (cf. Eq.(\ref{kappa_sh})) one thus has
\begin{equation}
\label{sum_p}
\sum_{\bf p}{\rm Tr}\left[ j^a_k \cG_n j_l\cG_{n+\nu}\right]
=
-\frac{2\pi}{e}\left[\sigma_{\rm sc}\right]^a_{kl}(\mu+{\rm i}\epsilon_n).
\end{equation}
This last result is derived in the appendix  more in detail. 
 
Eq.(\ref{eq6}) only takes into account the so-called bare bubble.  Vertex corrections
\cite{raimondi2001,schwab2002} will be considered later.
According to Eq.~(\ref{sum_p}) we now have
\begin{equation}
\label{eq7}
\sum_{\bf p}{\rm Tr}\left[ j^z_y \cG_n j_x\cG_{n+\nu}\right]
=
-\frac{2\pi}{e}\sigma^{sH}(\mu+{\rm i}\epsilon_n),
\end{equation}
with $\sigma^{sH}(\mu)$ the static spin-Hall conductivity from Ref.~\onlinecite{raimondi2005}.
The thermo-spin Hall conductivity therefore reads
\begin{equation}
\label{eq8}
{\rm N}^{sH}T=-\lim_{\Omega \rightarrow 0}\left[\frac{2\pi T}{e\Omega_{\nu}} \sum_{n =-\nu}^{-1}
{\rm i} \epsilon_{n+\nu/2}  \ \sigma^{sH}(\mu+{\rm i}\epsilon_n)\right]_{{\rm i} \Omega_{\nu}\rightarrow \Omega^R},
\end{equation}
and, after expanding in $\epsilon_n$ as shown in the appendix,  yields 
\begin{equation}
\label{eq9}
{\rm N}^{sH}|_{\rm bare}= -\frac{\pi^2 T}{3} \frac{m\alpha^2 \tau^2}{\pi}.
\end{equation}
To connect this result with that of Ref.~\onlinecite{ma2010},
in which ${\rm N}^{sH}T$ is computed in the clean limit, $\tau\rightarrow \infty$,
we rewrite Eq.~(\ref{cGrashba})
\begin{equation}
\label{eq10}
\cG_{n,\pm} = \left[\mu-\frac{p^2}{2m}\mp\alpha p+{\rm i} \ {\rm sgn}(\epsilon_n )\left( \frac{1}{2\tau}+|\epsilon_n|\right)\right]^{-1},
\end{equation}
and  note that, as long as the temperature is finite, in the $\tau\rightarrow \infty$ limit
the poles are $2\pi T$ away form the real axis.
Thus the effective replacement $1/2\tau \rightarrow \pi T$ in Eq.(\ref{eq9})
yields the clean limit result
\begin{equation}
\label{eq11}
{\rm N}^{sH}|_{\rm clean}= - \frac{m\alpha^2 }{12\pi T},
\end{equation}
in agreement with Ref.~\onlinecite{ma2010}.
Let us now discuss the vertex corrections.
Taking them into account corresponds to sending $j^z_y\rightarrow J^z_y$,
$j_x\rightarrow J_x$ and $j^h_x\rightarrow J^h_x$. At the level of the Born approximation
either vertex could be renormalized: the bubble with $J^z_y$ and $j^h_x$ or that with  $j^z_y$ and $J^h_x$
are equivalent.  Moreover, since we neglect inelastic processes, $J^h_x={\rm i}\epsilon_{n+\nu/2} J_x$.
For the Rashba case it is known that $J_x=0$, i.e. $\sigma^{sH}=0$, and thus we immediately obtain
\begin{equation}
\label{kappa_rashba}
{\rm N}^{sH}|_{\rm dressed} = 0.
\end{equation}
However, notice that Eq.~(\ref{eq8}) holds for any form of the SO interaction term $H_{\rm so}$,
no matter whether of intrinsic or extrinsic nature.
Therefore, once the spin-Hall conductivity $\sigma^{sH}$ of a given system is known,
its thermo-spin Hall conductivity ${\rm N}^{sH}$ will follow at once.
Even more generally, from the Matsubara formulation, Eqs.~(\ref{kappa_sh})-(\ref{sum_p}),
we conclude that the spin-heat response of a disordered, SO coupled Fermi gas
in the metallic regime is completely determined by its spin-charge response.
This result holds in 2D and 3D, in the presence of arbitrary elastic scattering processes,
possibly spin-dependent, and beyond the Born approximation, i.e. it has the same range
of applicability of the Wiedemann-Franz law discussed in Ref.~\onlinecite{niven2005}.
This is the first main result of our work, which, after a Sommerfeld expansion,
can be written in the very simple form
\begin{equation}
\label{kappa_sigma}
{\rm N}_{\rm sh} = -e{\cal L} T \sigma'_{\rm sc}(\mu).
\end{equation}
In other words Mott's formula for the electric thermopower $S=-e{\cal L}T\sigma'/\sigma$
has its symmetric spin equivalent
\begin{equation}
\label{spin_mott}
S_s = -e{\cal L}T\sigma'_{sc}/\sigma_{sc}.
\end{equation}
Whether a direct relation between $S_s$ and $S$ exists is however not obvious
and will be one of our next concerns.


\section{Spin Nernst effect and spin thermopower in electron and hole gases} 
\label{sec_appl}

Specializing our treatment to some specific systems, we now have a two-fold aim: 
(i) to look for the possibility of efficient heath-to-spin conversion, $S_s\gg1$; 
(ii) to establish a relation, if any, between $S_s$ and $S$.

\begin{figure}
\includegraphics[width=.47\textwidth]{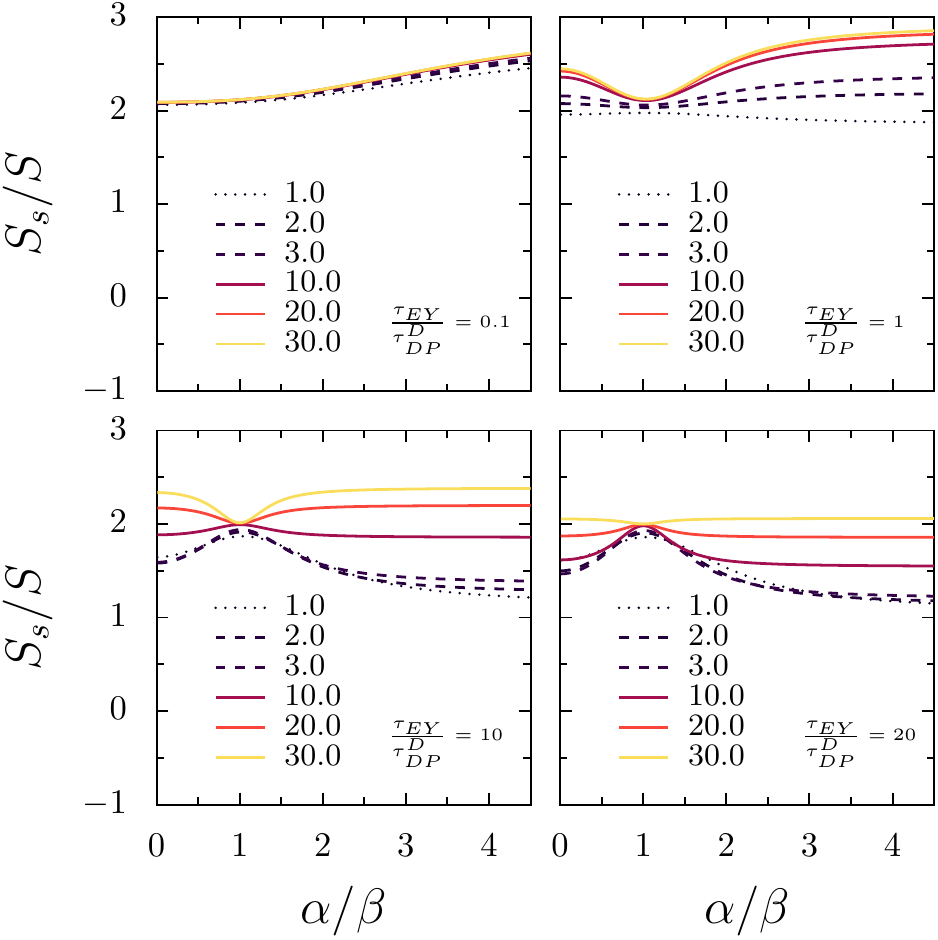}
\caption{The spin thermopower $S_s$ of a disordered 2D-electron gas
with numerous competing SO mechanisms.  Typical values
for GaAs quantum wells are: mobility $\mu=10^4{\rm cm}^2/{\rm Vs}$, density $n=10^{12}{\rm cm}^{-2}$,
effective extrinsic wavelength $\lambda_0=4.7 \times 10^{-8}{\rm cm}$,
Dresselhaus coupling constant $\hbar\beta=10^{-12}{\rm eVm}$.
There follows $\gamma_{ss}\gg\gamma_{intr},\gamma_{sj},\;\tau_{EY}\gg\tau_{DP}^D$.
The Rashba coupling constant can be modulated by the gate potential \cite{giglberger2007,yu2012}.
Each panel shows the ratio $S_s/S$ as a function of the ratio $\alpha/\beta$
for a given Elliot-Yafet scattering strength, strong to weak from top left to bottom right
-- panel 3 corresponds to standard GaAs.
Magnetic scattering is strongest for the dotted curve, $\tau_{sf}/\tau_{DP}^D=1$,
and strong (weak) for the dashed (solid) curves, $\tau_{sf}/\tau_{DP}^D=2,3\,(10,20,30)$.
}
\label{fig}
\end{figure}

With this in mind, let us now take $H_{\rm so}$ to be linear in momentum,
in which case the spin continuity equations assume a particularly simple form.
This allows one to easily draw a set of more specific conclusions concerning the thermo-spin response
of the 2D Fermi gas, in particular regarding the interplay between different SO and scattering mechanisms.
To be explicit we take once more the disordered Rashba model as the initial example,
and consider the presence of extrinsic SO mechanisms and (white noise) magnetic impurities.
That is, we add to the Hamiltonian (\ref{eq1}) the terms
\begin{equation}
H_{\rm extr} = - \frac{\lambda_0^2}{4}{\boldsymbol\sigma}\times\nabla V({\bf x})\cdot{\bf p},
\end{equation}
with $\lambda_0$ an effective Compton wavelength, and
\begin{equation}
V_{\rm m}(\bf x) = \sum_i\,{\bf B}\cdot{\boldsymbol\sigma}\delta({\bf x}-{\bf R}_i),
\end{equation}
where ${\bf B}$ is a random (white noise) magnetic field.
The latter is handled in the Born approximation,
$\langle V_{\rm m}({\bf x})V_{\rm m}({\bf x}')\rangle=[3(2\pi N_0\tau_{\rm sf})]^{-1}\delta({\bf x}-{\bf x}')$,
with $\tau_{\rm sf}$ the spin-flip time \cite{raimondi2012,gorini2008}.
The $s^y$ continuity equation reads
\begin{equation}
\label{continuity1}
\partial_t s^y + \nabla\cdot{\bf j}^y = -2m\alpha  j^z_y
-\left(
\frac{4}{3\tau_{\rm sf}}+\frac{1}{\tau_{\rm EY}}
\right)s^y,
\end{equation}
with $\tau_{\rm EY}=\tau(\lambda_0 p_F/2)^{-4}$ the Elliot-Yafet spin-relaxation time due to $H_{\rm extr}$.
Assuming a homogeneous electric field applied in the $x$-direction,
the spin current $j^z_y$ in the diffusive regime is given by
\begin{equation}
\label{current1}
j^z_y = 2m\alpha D s_y -\gamma\sigma E_x,
\end{equation}
where $\gamma=\gamma_{\rm intr}+\gamma_{\rm sj}+\gamma_{\rm ss}$ is the SO coupling constant due to
intrinsic and extrinsic mechanisms, with $\gamma_{\rm intr}=-m\alpha^2\tau$,
$\gamma_{\rm sj}=(\lambda_0/2)^2 m/\tau$ the side-jump contribution and
$\gamma_{\rm ss}=−(\lambda_0 p_F/4)^2(2\pi N_0 v_0)$ the skew-scattering one.
$v_0$ is the scattering amplitude (see Refs.~\onlinecite{raimondi2009,raimondi2010} for details).
In a homogeneous bulk in steady state the spin-Hall conductivity is easily computed,
$\sigma^{sH} = [1/(1+\zeta)]\gamma\sigma$,
where $\zeta\equiv\tau_{\rm s}/\tau_{DP}$, with $1/\tau_{DP}=(2m\alpha)^2D$
the Dyakonov-Perel spin-relaxation rate, and $1/\tau_{\rm s}\equiv4/(3\tau_{\rm sf})+1/\tau_{\rm EY}$.
Via Eq.~(\ref{kappa_sigma}) one concludes
\begin{eqnarray}
\label{sigma'_sH}
\sigma'^{sH} &=& \left[\frac{\sigma'}{\sigma}+\frac{\gamma'}{\gamma}-
\frac{\zeta'}{1+\zeta}\right]\sigma^{sH},
\\
\label{kappa_sH}
S_s&=& -e{\cal L} T  \left[\frac{\sigma'}{\sigma}+\frac{\gamma'}{\gamma}-
\frac{\zeta'}{1+\zeta}\right],
\end{eqnarray}
with the spin Hall thermopower $S_s={\rm N}^{sH}/\sigma^{sH}$.
In the above, primed quantities are derivatives with respect to the chemical potential $\mu$.
Notice that the simple phenomenological argument of the introduction overlooks
the $\mu$-dependency of $\gamma$: the conclusion $S_s=S$ holds only for an energy-independent $\gamma$.
Both $\sigma^{sH}$ and ${\rm N}^{sH}$ depend on the ratio 
between $\tau_{DP}$ and $\tau_{\rm s}$ and are in principle tunable, either by varying the doping, 
which affects $\tau_{\rm s}$, or by modulating $\alpha$ by varying the gate potential.
Let us consider some interesting cases using Eq.~(\ref{sigma'_sH}) and Eq.~(\ref{kappa_sH}).
When only Rashba SO and magnetic impurities are present, we have $\tau_{\rm s} = 3\tau_{\rm sf}/4$ and $\gamma=\gamma_{\rm int}$.
By evaluating  the various derivatives we obtain
$\gamma'=0,\,\zeta'=\zeta/\mu,\, \sigma'=\sigma/\mu$,
which gives us the spin thermopower
\begin{equation}
\label{spin_mag}
S_s= -e {\cal L}T \frac{\sigma'}{\sigma}\frac{1}{1+\zeta}.
\end{equation}
When SO from impurities is present, too, the terms $\gamma'/\gamma,\,\zeta'/\zeta$
in Eq.~\eqref{sigma'_sH} are modified, leading to
\begin{equation}
\label{spin_extm}
S_s= -e {\cal L}T \frac{\sigma'}{\sigma}
\left[1+\frac{\gamma_{\rm ss}}{\gamma}-\frac{\zeta}{1+\zeta}\left(1-\frac{2\tau_{\rm s}}{\tau_{\rm EY}}\right)\right].
\end{equation}
The results so far obtained can be generalized to include the effects 
of the linear-in-momentum Dresselhaus SO term
described by the Hamiltonian
\begin{equation}
\label{dresselhaus}
H_{so}=\beta \left( p_x \sigma^x -p_y\sigma^y\right).
\end{equation}
It suffices to replace in the above $\gamma_{\rm intr}=-m\tau(\alpha^2-\beta^2),\,
1/\tau_{DP}=(2m)^2\left(\alpha^2+\beta^2\right)D\equiv 1/\tau_{DP}^R+1/\tau_{DP}^D$
and
\begin{equation}
\zeta=\frac{\tau_{\rm s}}{\tau_{DP}}-4
\frac{\tau_{\rm s}^2/(\tau_{DP}^R\tau_{DP}^D)}{\tau_{\rm s}/\tau_{DP}+1}.
\end{equation}
Derivatives are trivial, but yield expressions too cumbersome to be conveniently written down.
The results are thus plotted in Fig.~\ref{fig}, and show the sensitivity
of the spin thermopower to the various physical parameters in play.
A modest modulation of the Rashba coupling constant could substantially modify $S_s$,
either enhancing or decreasing it depending on the system’s characteristics -- we considered ratios $\alpha/\beta$  well within 
current experimental capabilities \cite{giglberger2007,yu2012}.  We will come back to this point in a moment.
Let us now consider our final example, a 2D hole gas as analyzed in Ref.~\onlinecite{hughes2006}.
The SO interaction is cubic in momentum
\begin{equation}
\label{eqH}
H_{so}=\alpha_H\sigma_x\left[p_y\left(3p^2_x -p^2_y\right)\right]+
\alpha_H\sigma_y\left[p_x\left(3p^2_y -p^2_x\right)\right],
\end{equation}
and the spin Hall conductivity reads \cite{hughes2006}
\begin{equation}
\label{hole}
\sigma^{sH}_{H} = -\frac{3\eta^2\left(4\eta^2-1\right)}{\left(4\eta^2+1\right)^2}\frac{1}{\mu\tau}\sigma,
\end{equation}
with $\eta=\alpha_H p_{F}^3 \tau$ \cite{footnote1}. Proceeding as before one gets
\begin{equation}
\label{spin_hole}
S_s= -e {\cal L}T \frac{\sigma'}{\sigma}\left[\frac{3\left(12\eta^2-1\right)}{\left(4\eta^2+1\right) 
\left(4\eta^2-1\right)}\right].
\end{equation}
All previous result can be cast in the simple form
\begin{equation}
\label{generalform}
S_s=S R_s,
\end{equation}
with $R_s$ a number which depends on the various competing SO mechanisms.
Eq.~\eqref{generalform}, which is our second main result, looks physically quite reasonable: in a metallic system
in which electrons (or holes) are the sole carriers of charge, spin and heat,
the heat-to-spin and heat-to-charge (particle) conversions are expected to be closely related.
The examples considered show however that $R_s>1$ could be easily achieved:
in standard GaAs samples with Rashba SO and extrinsic mechanisms 
one may estimate $R_{so}\sim 3$ \cite{raimondi2009},
and the same value is obtained in a two-dimensional hole gas
with purely cubic Rashba SO in the diffusive regime ($\eta \ll 1$).
If Dresselhaus SO is also taken into account, similar values could be achieved, as shown in Fig.~\ref{fig}.
This suggests that metallic systems, typically characterized by low
thermoelectric efficiencies, could be much more efficient in heat-to-spin conversion
and therefore play a front role in spin caloritronics.
Of course, whether substantially higher $R_s$ values can be reached in different systems, e.g. in transition metals
which already show a giant spin Hall response \cite{seki2008, kontani2009},
or more exotic ones such as $p$-doped graphene \cite{tokatly2010}
or topological insulators like HgTe \cite{koenig2007}, is an open and relevant question. 
Indeed, it would be interesting to establish whether it is always possible,
within the regime in which the general expression \eqref{kappa_sigma} holds, 
to find such a simple connection between $S_s$ and $S$.
We therefore believe it desirable to experimentally test Eq.~\eqref{generalform}.
This could be done rather straightforwardly in a setup like the one employed to first observe
the spin Hall effect \cite{kato2004}: at low temperatures, the spin accumulation at the side edges of a two-dimensional
Fermi gas could be optically measured first in response to a longitudinally applied bias,
and then to a small temperature gradient along the same direction.
All-electrical measurement schemes based on H-bar geometries,
exchanging again the applied bias with a temperature difference, would also be interesting
though probably more delicate: in this case a temperature gradient along the side leg
of the H-bar should be avoided or its effects compensated.
Finally, it is well known that Mott's formula can be heavily affected by inelastic processes.
Though the latter are beyond the scope of the present work, it would be interesting 
to study their effects on $S_s$ and see whether any similarities between
electric and spin thermopower exist also in their presence or not.


\section{Conclusions}
\label{sec_conclusions}

In conclusion, we have studied coupled spin and thermal transport in a disordered and SO coupled Fermi gas,
and shown the existence of a general expression for the spin thermopower $S_s$
with the same structure and an identical range of validity of Mott's formula
for the electric thermopower $S$.  Finally, we have derived a simple and physically
transparent relation connecting the two quantities 
which could be experimentally tested and suggests that metallic systems
could be much more efficient in heat-to-spin than in heat-to-charge conversion. 

We acknowledge financial support from the EU through Grant. No. PITN-GA-2009-234970 and from the German Research Foundation DFG (TRR80).
CG acknowledges the hospitality of the IPCMS, Strasbourg, where part of this work was done, and thanks G.-L. Ingold for PyX support.


\appendix
\section{Derivation of Eqs.(\ref{sum_p}) and (\ref{eq9})}
\label{app_matsubara}
By defining
\begin{equation}
\label{F}
F({\rm i} \epsilon_n,{\rm i} \Omega_\nu)
=
\sum_{\bf p} {\rm Tr}\left[ j^a_k \cG_n j_l\cG_{n+\nu}\right],
\end{equation}
we write the spin-heat and spin-charge responses as
\begin{equation}
\label{sigma}
\sigma_{\rm sc}
=
\lim_{\Omega \rightarrow 0}
\left\{\frac{(-e) T}{\Omega_\nu}
\sum_{\epsilon_n}F({\rm i} \epsilon_n,{\rm i} \Omega_\nu)
\right\}_{{\rm i}\Omega_{\nu}\rightarrow \Omega^R},
\end{equation}
\begin{equation}
\label{n}
N_{\rm sh}
=
\lim_{\Omega \rightarrow 0}
\left\{\frac{1}{\Omega_\nu}
\sum_{\epsilon_n}{\rm i} \epsilon_{n+\nu/2}F({\rm i} \epsilon_n,{\rm i} \Omega_\nu)
\right\}_{{\rm i}\Omega_{\nu}\rightarrow \Omega^R}.
\end{equation} 
As mentioned in the main text, 
the momentum integral 
yields a non-zero result only if the frequencies
$\epsilon_n+\Omega_{\nu}$ and $\epsilon_n$ have opposite signs, which means that
$\epsilon_n$ is restricted to the range $-\Omega_{\nu} < \epsilon_n <0$.
Since the external frequency is going to zero, so will ${\rm i}\epsilon_n$, 
enabling one to expand $F$ in powers of ${\rm i}\epsilon_n$
\begin{equation}
\label{F1}
F({\rm i} \epsilon_n,{\rm i} \Omega_\nu)
=
F(0,{\rm i} \Omega_\nu)+{\rm i} \epsilon_n \frac{\partial{F}}{\partial{{\rm i} \epsilon_n}}(0,{\rm i} \Omega_\nu)+\dots.
\end{equation}
Replacing this expansion in Eq.\eqref{sigma} we have:
\begin{equation}
\label{sigma1}
\sigma_{\rm sc}
=
\lim_{\Omega \rightarrow 0}
\left\{\frac{e T}{\Omega_\nu}
 \sum_{n =-\nu}^{-1}F(0,{\rm i} \Omega_\nu)+{\rm i} \epsilon_n \frac{\partial{F}}{\partial{{\rm i} \epsilon_n}}(0,{\rm i} \Omega_\nu)+....
\right\}_{{\rm i}\Omega_{\nu}\rightarrow \Omega^R}
\end{equation}
The first term of the sum is linear in $\Omega_\nu$, 
so when divided by $\Omega_\nu$ in the zero-frequency limit it yields a non-zero contribution. 
The other terms of the sum, being at least quadratic in $\Omega_{\nu}$, clearly do not contribute.
There follows
\begin{equation}
\label{sigma2}
\sigma_{\rm sc}
=
-\frac{e}{2 \pi} F(0,0).
\end{equation}
This is enough to prove Eq.(\ref{sum_p}). To prove Eq.(\ref{eq9}), we
expand Eq.(\ref{n}) in ${\rm i}\epsilon_n$ and note that the zero order term of the sum vanishes
since
\begin{equation}
\sum_{-\Omega_{\nu}<\epsilon_n <0}  \left({\rm i}\epsilon_n +\frac{{\rm i}\Omega_{\nu}}{2}\right) =0.
\end{equation}
By noticing that
\begin{equation}
\sum_{-\Omega_{\nu}<\epsilon_n <0}  \left({\rm i}\epsilon_n +\frac{{\rm i}\Omega_{\nu}}{2}\right)
{\rm i}\epsilon_n =\frac{\pi^2T^2}{3}\nu (1-\nu^2),
\end{equation}
the only term  contributing linearly in $\Omega_\nu$ is the
first order one. 
This leads to
\begin{equation}
\label{n1}
{\rm N}_{\rm sh} = -e{\cal L} T F'(0,0),
\end{equation}
with ${\cal L}$ the Lorenz number and  $F'=\frac{\partial{F}}{\partial{{\rm i} \epsilon_n}}$.
The last step in proving Eq.(\ref{eq9}) of the main text is the observation that the
function $F$ of Eq.(\ref{F}) depends on $\epsilon_n$ through the combination
${\rm i} \epsilon_n+\mu$, as it is evident from the expression of the Green functions in the
restricted frequency range  $-\Omega_{\nu} < \epsilon_n <0$
\begin{equation}
\label{cGrashba3}
\cG_{n} = \left[{\rm i}\epsilon_n+\mu-\frac{{\rm i}}{2\tau}-H_{SO}\right]^{-1}
\end{equation}
\begin{equation}
\label{cGrashba4}
\cG_{n+\nu} = \left[{\rm i}(\epsilon_n+\Omega_{\nu})+\mu+\frac{{\rm i}}{2\tau}-H_{SO}\right]^{-1},
\end{equation}
where we have left unspecified the spin-orbit Hamiltonian for the sake of generality.


\begin{thebibliography}{999}
\bibitem{chester1961} G. V. Chester and Thellung, Proc. Phys. Soc. London {\bf 77}, 1005 (1961).
\bibitem{jonson1980} M. Jonson and G. D. Mahan, Phys. Rev. B {\bf 21}, 4223 (1980).
\bibitem{langer1962} J. S. Langer, Phys. Rev. B {\bf 128}, 110 (1962).
\bibitem{castellani1987} C. Castellani, C. Di Castro, G. Kotliar, P. A. Lee, and G. Strinati, Phys. Rev. Lett. {\bf 59}, 477 (1987).
\bibitem{livanov1991} D. V. Livanov, M. Y. Reizer, and A. V. Sergeev, Zh. Eksp. Teor. Fiz. {\bf 99}, 1230 (1991) [Sov. Phys. JETP {\bf 72}, 760 (1991)].
\bibitem{raimondi2004} R. Raimondi, G. Savona, P. Schwab, T. L\"uck, Phys. Rev. B {\bf 70},   155109 (2004).
\bibitem{niven2005} D. R. Niven and R. A. Smith, Phys. Rev. B {\bf 71}, 035106 (2005).
\bibitem{catelani2005} G. Catelani and I. L. Aleiner, Zh. Eksp. Teor. Fiz. {\bf 127}, 372 (2005) [Sov. Phys. JETP {\bf 100}, 331 (2005)].
\bibitem{michaeli2009} K. Michaeli  and A. M. Finkel�stein, Phys. Rev. B {\bf 80}, 115111 (2009).
\bibitem{sondheimer1948} E. H. Sondheimer, Proc. R. Soc. Lond. A  {\bf 193}, 484 (1948).
\bibitem{zutic2004} I. $\check{\rm Z}$uti\'c, J.Fabian, and S. D. Sarma,  Rev. Mod. Phys. {\bf 76}, 323 (2004).
\bibitem{awschalom2007} D. D. Awschalom and M. E. Flatt\'e,  Nat.  Phys. {\bf 3}, 153 (2007).
\bibitem{bauer2012} G. E. W. Bauer, E. Saitoh, and B. J. van Wees, Nat. Mat. {\bf 11}, 391 (2012).
\bibitem{dyakonov1971}  M. I. Dyakonov and V. I. Perel, Phys. Lett. A {\bf 35},  459, (1971).
\bibitem{hirsch1999} J. E. Hirsch, Phys. Rev. Lett. {\bf 83}, 1834 (1999).
\bibitem{murakami2003} S. Murakami, N. Nagaosa, and S.-C. Zhang, Science {\bf 301}, 1348
(2003).
\bibitem{sinova2004}  J. Sinova, D. Culcer, Q. Niu, N. A. Sinitsyn, T. Jungwirth, and A.
H. MacDonald, Phys. Rev. Lett. {\bf 92}, 126603 (2004).
\bibitem{vignale2010} G.Vignale, J.Supercond. Nov. Magn. {\bf 23}, 3 (2010).
\bibitem{wang2010} C.M. Wang, M.Q. Pang, Solid State Communications {\bf 150}, 1509 (2010).
\bibitem{ma2010} Z. Ma, Solid State Communications, {\bf 150}, 510 (2010).
\bibitem{nunner2011} T. S. Nunner and F. von Oppen, Phys. Rev. B {\bf 84}, 020405(R) (2011).
\bibitem{slachter2011} A. Slachter, F. L. Bakker, and Bart Jan van Wees, Phys. Rev. B {\bf 84}, 174408 (2011).
\bibitem{scharf2011} B. Scharf, A. Matos-Abiague, I. $\check{\rm Z}$uti\'c, and J. Fabian, arxiv:1112.1808v2.
\bibitem{uchida2008} K. Uchida, S. Takahashi, K. Harii, J. Ieda, W. Koshibae, K. Ando, S. Maekawa and E. Saitoh,
Nat. Lett. {\bf 455}, 778 (2008).
\bibitem{uchida2010} K. Uchida, J. Xiao, H. Adachi, J. Ohe, S. Takahashi, J. Ieda, T. Ota, Y. Kajiwara, H. Umezawa, H. Kawai, G.E.W. Bauer,
S. Maekawa and E. Saitoh,  Nat. Mater. {\bf 9}, 894 (2010).
\bibitem{jaworski2010} C. M. Jaworski, J. Yang, S. Mack, D. D. Awschalom, J. P. Heremans and R. C. Myers,
Nat. Mat. {\bf 9}, 898 (2010).
\bibitem{jaworski2011} C. M. Jaworski, J. Yang, S. Mack, D. D. Awschalom, R. C. Myers and J. P. Heremans,
Phys. Rev. Lett. {\bf 106}, 186601 (2011).
\bibitem{uchida2012} K. Uchida, T. Ota, H. Adachi, J. Xiao, T. Nonaka, Y. Kajiwara, G. E. W. Bauer, S. Maekawa and E. Saitoh,
J. Appl. Phys. {\bf 111}, 103903 (2012). 
\bibitem{adachi2012} H. Adachi K. Uchida, E. Saitoh and S. Maekawa, arXiv:1209.6407 (2012).
\bibitem{schwab2010} P. Schwab, R. Raimondi, and C. Gorini, EPL {\bf 90}, 67004 (2010).
\bibitem{dyakonov2007} M. I. Dyakonov, Phys. Rev. Lett.  {\bf 99}, 126601 (2007).
\bibitem{aschcroftbook} N. Aschcroft, and W. Mermin, {\cal Solid State Physics}, 
Harcourt College Publishers, Orlando FL  (1976).
\bibitem{raimondi2001} R. Raimondi, M. Leadbeater, P. Schwab, E. Caroti and C. Castellani, Phys. Rev. B {\bf 64}, 235110 (2001).
\bibitem{schwab2002}  P. Schwab and R. Raimondi, Eur. Phys. J. B {\bf 25}, 483 (2002).
\bibitem{raimondi2005} R. Raimondi and P. Schwab, Phys. Rev. B {\bf 71}, 033311 (2005).
\bibitem{gorini2008} C. Gorini, P. Schwab, and M. Dzierzawa and R. Raimondi, Phys. Rev. B {\bf 78}, 125327 (2008).
\bibitem{raimondi2012} R. Raimondi, P. Schwab, C. Gorini, and G. Vignale,
Ann. Phys. (Berlin) {\bf 524}, 153 (2012).
\bibitem{raimondi2009} R. Raimondi and P. Schwab, EPL  {\bf 87}, 37008 (2009).
\bibitem{raimondi2010} R. Raimondi and P. Schwab, Physica E {\bf  42}, 952 (2010).
\bibitem{giglberger2007} S. Giglberger, L. E. Golub, V. V. Bel'kov, S. N. Danilov, D. Schuh, C. Gerl, F. Rohlfing,
J. Stahl, W. Wegscheider, D. Weiss, W. Prettl  and S. D. Ganichev, Phys. Rev. B {\bf 75}, 035327 (2007).
\bibitem{yu2012} J. L. Yu, Y. H. Chen, Y. Liu, C. Y. Jiang, H. Ma and L. P. Zhu, Appl. Phys. Lett. {\bf 100} 152110 (2012).
\bibitem{hughes2006} T. L. Hughes, Y. B. Bazaliy, and B. A. Bernevig, Phys. Rev. B {\bf 74}, 193316 (2006).
\bibitem{footnote1} The parameter $\eta$ corresponds to what the authors of Ref.~[\onlinecite{hughes2006}] call $\zeta$.
\bibitem{seki2008} T. Seki, Y. Hasegawa, S. Mitani, S. Takahashi, H. Inamura, S. Maekawa, J. Nitta,
and K. Takanashi, Nat. Mat. {\bf 7}, 125 (2008).
\bibitem{kontani2009} H. Kontani, T. Tanaka, D. S. Hirashima, K. Yamada, and J. Inoue,
 Phys. Rev. Lett. {\bf 102}, 016601 (2009).
\bibitem{tokatly2010} I. V. Tokatly, Phys. Rev. B {\bf 82}, 161404(R) (2010).
\bibitem{koenig2007} M. K\"{o}nig, S. Wiedmann, C. Br\"{u}ne, A. Roth, H. Buhmann, L. W. Molenkamp, 
X.-L. Qi, S.-C. Zhang, Science {\bf 318}, 766 (2007).
\bibitem{kato2004} Y. K. Kato, R. C. Myers, A. C. Gossard, D. D. Awschalom, Science {\bf 306}, 1910 (2004).
\end{thebibliography}
\end{document}